\documentclass[]{aa}
\usepackage{graphics}
\begin{document}
\thesaurus{08(08.16.6; 02.18.5)}
\title{Pulsar spectra of radio emission}

\author{Olaf Maron\inst{1}\and Jaros\l aw Kijak\inst{1}\and Michael Kramer\inst{2,3}\and Richard Wielebinski\inst{2}}

\offprints{Olaf Maron, e-mail: olaf@astro.ca.wsp.zgora.pl}
\institute{J. Kepler Astronomical Center, Pedagogical University, Lubuska 2, PL-65-265 Zielona G\'ora, Poland
\and Max-Planck-Institut f\"ur Radioastronomie, Auf dem H\"ugel 69, D-53121 Bonn, Germany
\and University of Manchester, Jodrell Bank Observatory, Macclesfield, Cheshire SK11 9DL, United Kingdom}

\date{Received 11 April 2000 / Accepted 7 September 2000}

\maketitle

\begin{abstract}

We have collected pulsar flux density observations and compiled spectra of 281
objects. The database of Lorimer et al. (\cite{lorimer}) has been extended to frequencies higher than 1.4 GHz and lower than 300 MHz. Our results show that above 100 MHz the spectra of the majority of pulsars can be described by a simple power law with average value of spectral index $<\alpha> = -1.8\pm0.2$. A rigorous analysis of spectral fitting revealed only about 5\% of spectra which can be modelled by the two power law. Thus, it seems that single power law is a rule and the two power law spectrum is a rather rare exception, of an unknown origin, to this rule. We have recognized a small number of pulsars with almost flat spectrum ($\alpha \geq -1.0$) in the wide frequency range (from 300 MHz to 20 GHz) as well as few pulsars with a turn-over at unusually high frequency ($\sim$~1~GHz).

\keywords{pulsars: general - radiation mechanisms: non-thermal}
\end{abstract}

\section{Introduction}

One of the main observables of pulsar emission is its flux density
${S_{\nu }}$, measured at a given central frequency $\nu $ of the
receiver bandwidth. Flux measurements are crucial for deriving, the so
called, pulsar luminosity function and therefore the birth rate of the
Galactic population of neutron stars. The first spectrum of a pulsar
taken at 5 frequencies was published by Robinson et
al. (\cite{robinson}). The flux density variations and spectra for
frequencies between 0.15 and 1.6 GHz were later reported for a number
of pulsars by McLean (\cite{mclean}). Spectra of 27 pulsars were
published by Sieber (\cite{sieber73}) who used pulse energy values
obtained with the 100-m and 25-m telescopes of the Max-Planck-Institute
for Radioastronomy (MPIfR) as well as other values published in the
literature. He was the first to show a turn-over behaviour at low frequencies around 100 MHz and a break
in spectrum at high frequencies (about 1 GHz). Sieber et al. (\cite{sieber75})
published pulse shapes and their energies for 35 pulsars at 2.7 
and 4.9 GHz, and for 7 pulsars at 10.7 GHz. Izvekova et
al. (\cite{izvekova}) and Slee et al. (\cite{slee}) presented an analysis
of flux measurements for pulsars at meter wavelengths ($\sim$~80
MHz). A compilation of spectra of 45 pulsars over a wide frequency
range was published for the first time by Malofeev et
al. (\cite{malofeev94}). Seiradakis et al. (\cite{seiradakis}) published a collection of high frequency data on pulsar profiles and
flux densities for 183 pulsars at 1.4 GHz, 46 pulsars at 4.85 GHz and
24 pulsars at 10.5 GHz. The catalogue of pulsar flux density
measurements for 281 pulsars at frequencies ranging from 0.4 GHz to
1.6 GHz was published by Lorimer et al. (\cite{lorimer}). The first
spectra for four millisecond pulsars were published by Foster et
al. (\cite{foster}). Kramer et al. (\cite{kramer98}) presented the
results of flux density measurements for 23 millisecond pulsars at
frequencies 1.4 and 1.7 GHz and an analysis of their spectra. Van Ommen
et al. (\cite{ommen}) presented polarimetric data together with flux
density measurements for a large number of southern pulsars at
frequencies of 800 MHz and 950 MHz. Most recently, Toscano et
al. (\cite{toscano}) have presented flux density measurements for
southern millisecond and slow pulsars at frequencies between 0.4 GHz
and 1.6 GHz, and Kramer et al.~(\cite{kramer99}) studied the emission properties
of millisecond pulsars up 4.85 GHz.
A first large sample of flux densities of weak pulsars at
4.85 GHz was published by Kijak et al. (\cite{kijak}).

It has been long believed that flux density spectra for most of the pulsars
have been well described by a simple power law $S \propto \nu ^{\alpha }$
with spectral index of $\sim {-1.6}$ (Sieber \cite{sieber73} and
Lorimer et al. \cite{lorimer}). However, a considerable fraction of pulsars
demonstrated spectra that required modelling by two power laws (Sieber
\cite{sieber73} and Malofeev et al. \cite{malofeev94}). These were
commonly called the broken-type spectra. In this paper we show that the two power law spectra are rare exception and that majority of pulsars can be modelled by a single power law. Our analysis shows that only 10\% of pulsars requires two power law spectra.

Recently, pulsar flux density measurements
have been extended to the mm-wavelengths region, which provided
information about this newly explored spectral region. These
measurements suggested that spectrum flattens out or even turns up at
very high frequencies (Kramer et al. \cite{kramer96}).  In this paper
we present a new and most complete compilation of spectra for 281
pulsars in the frequency range from 39 MHz to 43 GHz. The flux density measurements for frequencies from 300 MHz to 1.4 GHz were done at Jodrell Bank (Lorimer et al. \cite{lorimer}). For frequencies equal or above 1.4
GHz we have utilized flux measurements made by different authors at
the Effelsberg Radiotelescope of the Max-Planck Institute for
Radioastronomy (Sieber et al. \cite{sieber75}, Bartel et
al. \cite{bartel}, Sieber \& Wielebinski \cite{sieber87}, Wielebinski
et al. \cite{wielebinski}, Seiradakis et al. \cite{seiradakis}, Kramer
et al. \cite{kramer96}, Hoensbroech et al. \cite{hoensbroech},
Kramer et al. \cite{kramer97}, Kijak et al. \cite{kijak}). We have
also reduced a significant amount of the unpublished data available at
the archives of the MPIfR. Most of these data is available throughout the
European Pulsar Network Database (Lorimer et al. \cite{lorimer98}). We
have also included the data published by Izvekova et
al. (\cite{izvekova}) and Malofeev et al. (\cite{malofeev2000})
containing observations made at low radio frequencies from 39 MHz to
102.5 MHz. Our own observations at 4.85 GHz made in Effelsberg in 1998
are also included.

\section{Observations and data reduction}

As a basis for our database we have taken the flux density
measurements published by Lorimer et al. (\cite{lorimer}). These
observations were made between July 1988 and October 1992 using the
76-m Lovell radio telescope at Jodrell Bank, at frequencies 408, 606,
925, 1408 and 1606 MHz.  Lorimer et al. (\cite{lorimer}) excluded from
their sample those pulsars which were too weak to obtain reliable flux
density measurements as well as the millisecond pulsars. We have
extended this database by observations made at higher frequencies by
different authors mentioned earlier or those unpublished but made
available at MPIfR archives in Bonn. All observations at frequency
range from 1.4 GHz to 43 GHz were made with the 100-m radio telescope
of the MPIfR at Effelsberg and are available in European Pulsar
Network Database (Lorimer et al. \cite{lorimer98}). Some values at 1.4
and 1.6 GHz were also published by Lorimer et al. (\cite{lorimer}). We
performed additional observations at 4.85 GHz of 43 very weak pulsars
in August 1998. We managed to detect 30 objects and those undetected
are listed in Table 1. The detection limit of $\rm S\sim 0.05~mJy$ for
the survey published by Kijak et al. (\cite{kijak}) is clearly visible
from this table. The observations published by Izvekova et
al. (\cite{izvekova}) and Malofeev et al. (\cite{malofeev2000}) were
performed at the Pushchino Radio Astronomical Observatory of the
Lebedev Physical Institute.

\begin{table}
\caption{Pulsars not detected at 4.85 GHz. The pulsar name, total observing time and upper limits $S_{\rm max}$ for the total flux density are listed.}
\begin{tabular}{llllll}
\hline
PSR B & Time & $\rm S_{max}$ & PSR B & Time & $\rm S_{max}$ \\
 & [min] & [mJy] &  & [min] & [mJy] \\
\hline
0621$-$04 & 30 & 0.02 & 1820-14 & 20 & 0.02\\
1246+22 & 50 & 0.01 & 1834-04 & 30 & 0.01\\
1534+12 & 20 & 0.01 & 1839-04 & 25 & 0.03\\
1600$-$27 & 25 & 0.01 & 1848+12 & 20 & 0.01\\
1811+40 & 20 & 0.01 & 2210+29 & 100 & 0.01\\
1813$-$17 & 25 & 0.01 & 2323+63 & 20 & 0.03\\
\hline
\end{tabular}
\end{table}
\subsection{Calibration procedure}

In order to calibrate the flux density of a pulsar using a system in
the Effelsberg Radio-observatory, a noise diode installed in every
receiver is switched synchronously with the pulse period. The energy
output for the noise diode is then compared with energy received from
the pulsar, since the first samples of an observed pulse profile
contain the calibration signal while the remaining samples contain the
pulse. The energy of the noise diode itself can be calibrated by
comparing its output to the flux density of a known continuum
sources. This pointing procedure is generally performed on well known
reliable flux calibrators, e.g. 3C123, 3C48, etc. From these pointing
observations a factor $f_{c}$ translating the height of the
calibration signal into flux units, is derived.  The energy of a pulse
is given as the integral beneath its waveform which in arbitrary units
yields
\begin{equation}
E=\sum\limits_{i\in {\rm pulse}}A_{i}\Delta t_{{\rm samp}}=\Delta t_{{\rm samp}}\sum\limits_{i\in {\rm pulse}}A_{i}
\end{equation}
where $A_i$ is the pulse amplitude measured in the phase bin $i$ and $\Delta t_{{\rm {samp}}}$ is the sampling time. Scaling the whole profile now in units of the height of the calibration signal measured in the same profile, $A_{{\rm cal}}$, we obtain
\begin{equation}
E=\Delta t_{{\rm samp}}\sum\limits_{i\in {\rm pulse}}\frac{A_{i}}{A_{{\rm {cal}}}}.
\end{equation}
Using the conversion factor $f_{c}$, the mean pulse energy is translated into proper units, i.e. $\rm J~m^{-2}Hz$, according to the following formula
\begin{equation}
E=f_{c}\cdot\frac{\Delta t_{{\rm samp}}}{A_{cal}}\sum\limits_{i\in {\rm pulse}}A_{i}.
\end{equation}
The mean flux represents the pulse energy averaged over a pulse period,
\begin{equation}
S_{{\rm mean}}=\frac{E}{P}
\end{equation}
Finally, assuming that $f_{c}$ converts the units of the calibration signal strength into mJy, and the sampling time $\Delta t_{{\rm samp}}$ is given in $\mu$s, the mean flux is obtained as (Kramer \cite{kramerPhD})
\begin{equation}
S_{{\rm mean}}=10^{-3}\cdot\frac{f_{c}}{P}\cdot\frac{\Delta t_{{\rm samp}}}{A_{cal}}\sum\limits_{i\in {\rm pulse}}A_{i}.
\end{equation}

\subsection{Error analysis}

Pulsars are generally known to be stable radio sources although the
measured flux density varies due to diffractive and refractive
scintillation effects (e.g.,~Stinebring \& Condon \cite{sc90}).  Interstellar scintillations are
caused by irregularities in the electron density of the interstellar
medium. The observed flux variations are frequency and distance
dependent and also depend on the observing set-up, i.e.~on the
relative width of observing and scintillation bandwidth
(e.g.,~Malofeev et al. \cite{malofeev96a}, Malofeev \cite{malofeev96}). Unless the
receiver bandwidth is significantly larger than the scintillation
bandwidth, which increases with frequency, strong variations in the
observed flux densities are to be expected. Usually, however, the
amplitude of scintillation decreases towards higher frequencies, so
that those data are less influenced by the scintillation
effects. Nevertheless, the question of intrinsic variations on very
short and very long time scales remains still open (cf.~Stinebring \&
Condon \cite{sc90}).  Assessing the situation is hampered by the fact that
many authors do not estimate the always present influence of
interstellar scintillations or do not quote error estimates at
all. Given the difference in the observing set-ups for given
observatories, a careful analysis is difficult.  We try to circumvent
this unpleasant situation by estimating errors for the pulsar flux
densities in our sample from published values, wherever available, and
standard deviations of the average of single measurements.  If only
one measurement was available, an error estimate could not be computed
although it may happen that form of the spectrum changes when new
measurements are added.

\subsection{Search technique for break frequency}

Since a robust theory of pulsar radio emission does not exist, the
true shape of pulsar spectra is still not known. A fair first attempt
is to model them by simple power laws.  Previous studies (e.g. Sieber
\cite{sieber73}, Malofeev et al. \cite{malofeev94}) showed however
that some pulsar spectra cannot apparently be described by this
simple approach. Usually, such a conclusion is reached after a visual
inspection of the data, i.e.~after a power law fit has been
done. However, if pulsar spectra are indeed more complicated, the
usual next step to fit a composite (or 'broken') power law is just
another approximation, where the undersampling of the spectrum in the
observed range of frequencies would place any fitted 'break' naturally
in the range of a few GHz, i.e. the range where they are indeed
usually observed as it was clearly pointed out by Thorsett
(\cite{thorsett}). Nevertheless, even if two power laws are just
another approximation to a 'true' spectrum, any need to fit a break in
order to describe the data adequately would represent a valuable hint
on the true nature of pulsar spectra. It is therefore very important
to search for such breaks in the spectra, while keeping just discussed limitations in mind. We believe, however, that one has
to be more quantitative when describing the need for fitting a two power laws 
rather than a simple power law. This is even more important in the
light of the latest results on millisecond pulsar spectra (Kramer et
al. \cite{kramer99}, Kuzmin \& Losovski \cite{kl00}), where no significant
break (not even a low frequency turn-over) has been found. Hence, we
adopted in this work the following approach: Firstly, we fitted a
simple power law to the flux density data, assuming that this
describes the data sufficiently.  We calculated a $\chi^2$ and the
probability $Q$ that a random $\chi^2$ exceeds this value for a given
number of degrees of freedom. These computed probabilities give a
quantitative measure for the goodness-of-fit of the model. If $Q$ is
very small then the apparent discrepancies are unlikely to be chance
fluctuations. Much more probably either the model is wrong, or the
measurement errors are larger than stated, or measurement errors might
not be normally distributed. Generally, one may accept models with $Q
\sim$ 0.001 (Press et al. \cite{recipees}). Secondly, we assumed that
a two power law had to be fitted to the data, using the following
rules:

\begin{equation} 
S(\nu) = \left \{
\begin{array}{r@{\quad:\quad}l}
 c_1 \nu^{\alpha_1} &  \nu \le \nu_b \\ 
 c_2 \nu^{\alpha_2} &  \nu > \nu_b~~~.
\end{array} \right.
\end{equation} 

Since the break frequency, $\nu_b$, is {\em a priori} unknown, we treated it as a free parameter and tried to minimize a corresponding
$\chi^2$ simultaneously over the whole parameter space of $c_1, c_2,
\alpha_1, \alpha_2$ and $\nu_b$. Due to the nature of the additional boundary
condition ($\nu\le\nu_b$ or $\nu>\nu_b$), we applied a Simplex
algorithm as described by Nelder \& Mead (\cite {nelder}). For the
resulting, minimized $\chi^2$ we then calculated again the probability
that a random $\chi^2$ is larger than the found value. A comparison of
the $\chi^2$-statistics for both cases was then used to judge whether
a break was truly significant or not. The fitting procedure was
performed on data in the frequency range from 400 MHz to 23 GHz. We
have not taken into account data corresponding to single measurements, as well as those at frequencies lower than 400 MHz, as they could represent a low
frequency turn-over which usually occurs at $\sim$~100 MHz. There is a
gap in data coverage between 100 and 300 MHz, but this should not
affect our analysis and conclusions.

\begin{figure}
\setlength{\unitlength}{1cm}
\begin{picture}(1,13)
\put(-1,-0.8){\includegraphics{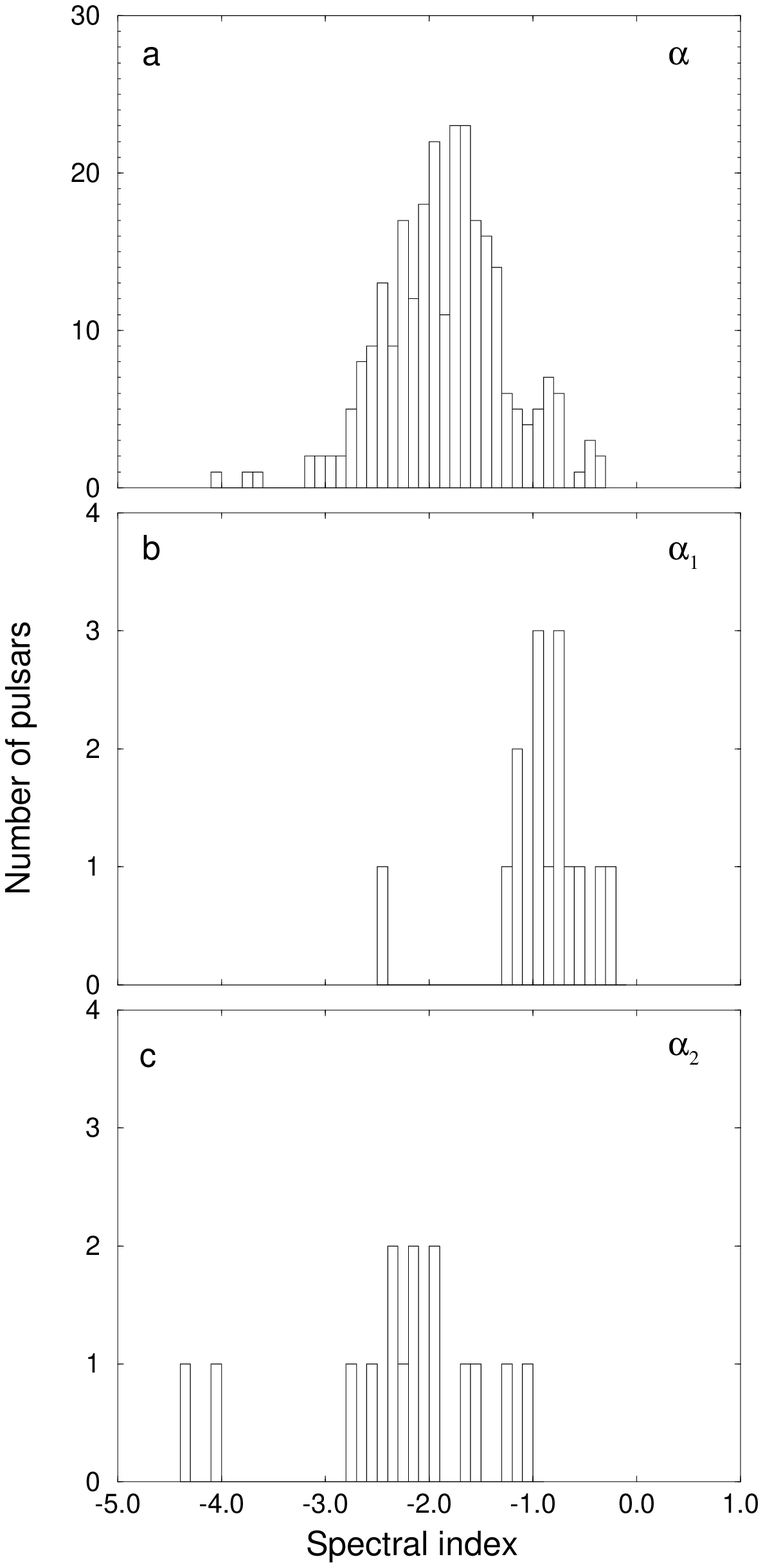}}
\end{picture}
\caption{{\bf a} The distribution of spectral index $\alpha$ for simple power law spectra,
{\bf b} The distribution of spectral index $\alpha_1$ for the low frequency part of two-power-law spectra,
{\bf c} The distribution of spectral index $\alpha_2$ for the high frequency part of two-power-law spectra.
}
\label{1}
\end{figure}

\begin{figure}
\setlength{\unitlength}{1cm}
\begin{picture}(1,13)
\put(-0.6,-1.2){\includegraphics{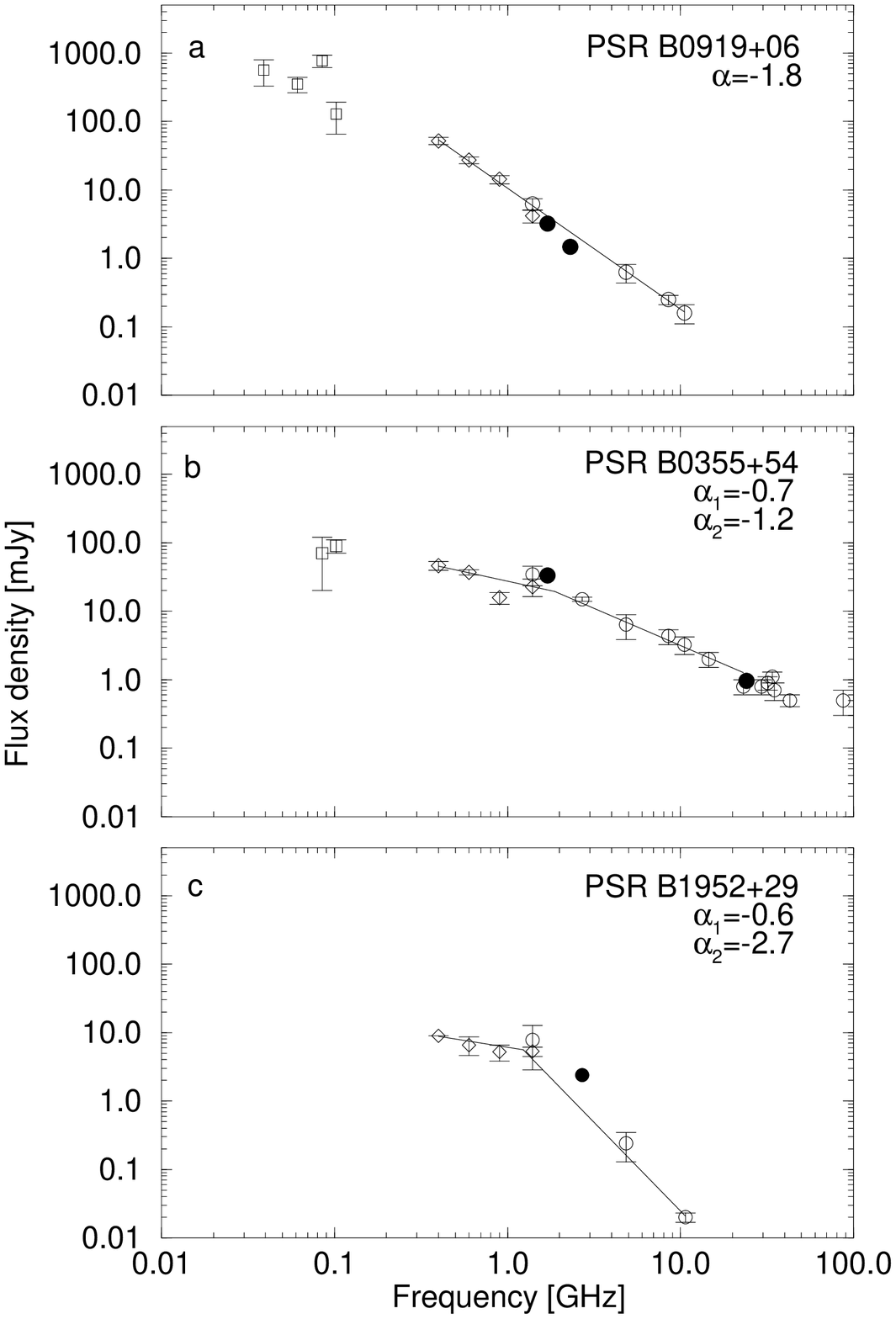}}
\end{picture}
\caption{{\bf a} Example of a typical spectrum with $\alpha=-1.8$,
{\bf b} Example of the two-power law spectrum with the smallest difference in slopes ($\alpha_1=-0.7$ and $\alpha_2=-1.2$),
{\bf c} Example of the two-power law spectrum with the largest difference in slopes ($\alpha_1=-0.6$ and $\alpha_2=-2.7$). Squares represent the measurements from Pushchino Radio Astronomical Observatory, diamonds represent measurements from Jodrell Bank and circles represent measurements from the Effelsberg Radiotelescope. Filled circles represent single measurements.
}
\label{2}
\end{figure}

\section{Results and discussion}

In this paper we obtained a large database of flux density
measurements over a wide range of frequencies, from 39 MHz to 43 GHz. Although measurements were made in three different
observatories, the data seem consistent. In particular,
the values obtained by
Lorimer et al. (\cite{lorimer}) and those from the Effelsberg
radio-telescope at the same frequency are comparable (e.g., PSR
B0740$-$28)\footnote{All figures with spectra of 281 are available at: http://astro.ca.wsp.zgora.pl/olaf/paper1}. We have calculated the spectral index for pulsars using the method described in Sect. 2.3. The results of
this analysis are listed in Table 2.

\setcounter{table}{1}
\begin{table*}
\caption{Spectral indices for 266 pulsars with simple power-law spectrum $S\sim \nu ^{\alpha }$  calculated for a given frequency range with error and a probability of the goodness-of-fit Q (see text). }
\begin{tabular}{lllllllllll}
\hline
PSR B & Freq. range & $\alpha$ & $\sigma_\alpha$ & Q& &PSR B & Freq. range & $\alpha$ & $\sigma_\alpha$ & Q \\
 & [GHz] & & & \\
\hline
0011+47&0.4 -- 4.9&-1.3&0.10&8.3E-01&&0943+10&0.4 -- 0.6&-3.7&0.36&\\
0031$-$07&0.4 -- 10.7&-1.4&0.11&4.1E-03&&0950+08&0.4 -- 10.6&-2.2&0.03&1.7E-02\\
0037+56&0.4 -- 4.8&-1.8&0.05&6.3E-03&&1010$-$23&0.4 -- 0.6&-1.9&&\\
0045+33&0.4 -- 1.4&-2.5&0.26&&&1016$-$16&0.4 -- 1.4&-1.7&0.28&9.1E-01\\
0052+51&0.4 -- 1.4&-0.7&0.14&6.1E-01&&1039$-$19&0.4 -- 1.4&-1.5&0.28&6.8E-01\\
0053+47&0.4 -- 4.9&-1.6&0.09&&&1112+50&0.4 -- 4.9&-1.7&0.11&6.0E-02\\
0059+65&0.4 -- 1.6&-1.6&0.13&1.4E-01&&1133+16&0.4 -- 32.0&-1.9&0.06&7.4E-02\\
0105+65&1.4 -- 1.4&-1.9&0.19&4.3E-01&&1254$-$10&0.4 -- 1.6&-1.8&0.16&6.4E-01\\
0105+68&0.4 -- 1.4&-1.8&0.22&&&1309$-$12&0.4 -- 1.4&-1.7&0.16&2.8E-01\\
0114+58&0.4 -- 1.4&-2.5&0.21&8.5E-01&&1322+83&0.4 -- 1.4&-1.6&0.30&8.7E-01\\
0138+59&0.4 -- 1.4&-1.9&0.16&7.0E-01&&1508+55&0.4 -- 4.9&-2.2&0.07&5.2E-02\\
0144+59&0.4 -- 14.6&-1.0&0.04&6.1E-04&&1530+27&0.4 -- 4.9&-1.4&0.10&1.3E-01\\
0148$-$06&0.4 -- 1.4&-2.7&0.58&4.6E-01&&1540$-$06&0.4 -- 4.9&-2.0&0.11&4.0E-02\\
0149$-$16&0.4 -- 1.4&-2.1&0.26&2.3E-01&&1541+09&0.4 -- 4.9&-2.6&0.04&2.7E-03\\
0153+39&0.4 -- 0.6&-2.2&&&&1552$-$23&0.4 -- 4.9&-1.8&0.08&1.1E-02\\
0154+61&0.4 -- 1.4&-0.9&0.12&1.4E-03&&1552$-$31&0.4 -- 1.4&-1.6&0.19&3.0E-04\\
0320+39&0.4 -- 1.4&-2.9&0.24&5.2E-01&&1600$-$27&0.4 -- 1.4&-1.7&0.13&4.0E-04\\
0329+54&1.4 -- 23.0&-2.2&0.03&7.5E-04&&1604$-$00&0.4 -- 4.9&-1.5&0.08&6.8E-01\\
0331+45&0.4 -- 1.4&-1.9&0.24&1.5E-01&&1607$-$13&0.4 -- 0.6&-2.1&0.45&\\
0339+53&0.4 -- 1.4&-2.2&0.28&2.5E-02&&1612+07&0.4 -- 1.4&-2.6&0.30&1.7E-01\\
0353+52&0.4 -- 1.4&-1.6&0.12&7.4E-01&&1612$-$29&0.4 -- 0.6&-0.8&0.98&\\
0402+61&0.4 -- 1.4&-1.4&0.08&1.4E-01&&1620$-$09&0.4 -- 4.9&-1.7&0.13&1.5E-01\\
0410+69&0.4 -- 1.4&-2.4&0.13&8.1E-04&&1633+24&0.4 -- 1.4&-2.4&0.31&\\
0447$-$12&0.4 -- 1.4&-2.0&0.11&1.4E-02&&1642$-$03&0.4 -- 10.6&-2.3&0.05&6.7E-01\\
0450+55&0.4 -- 4.9&-1.5&0.04&6.0E-02&&1648$-$17&0.4 -- 1.4&-2.5&0.26&5.9E-02\\
0450$-$18&0.4 -- 4.9&-2.0&0.05&2.9E-08&&1649$-$23&0.4 -- 1.4&-1.7&0.09&5.6E-01\\
0458+46&0.4 -- 1.4&-1.3&0.05&1.2E-05&&1657$-$13&0.4 -- 0.6&-1.7&0.36&\\
0523+11&0.4 -- 1.4&-2.0&0.06&1.4E-01&&1700$-$18&0.4 -- 1.4&-1.9&0.23&4.8E-05\\
0525+21&0.4 -- 4.9&-1.5&0.12&4.4E-01&&1700$-$32&0.4 -- 0.6&-3.1&0.27&\\
0531+21&0.4 -- 1.4&-3.1&0.18&5.7E-01&&1702$-$19&0.4 -- 4.9&-1.3&0.05&5.3E-01\\
J0538+2817&1.4 -- 4.9&-1.2&0.57&&&1706$-$16&0.4 -- 32.0&-1.5&0.04&1.1E-01\\
0559$-$05&0.4 -- 4.9&-1.7&0.04&7.6E-01&&1709$-$15&0.4 -- 4.9&-1.7&0.06&8.7E-01\\
0609+37&0.4 -- 1.4&-1.5&0.25&3.5E-01&&1714$-$34&0.4 -- 1.4&-2.6&0.34&\\
0611+22&0.4 -- 2.7&-2.1&0.04&8.5E-01&&1717$-$16&0.4 -- 4.9&-2.2&0.05&5.7E-01\\
0621$-$04&0.4 -- 1.4&-0.4&0.29&5.0E-01&&1717$-$29&0.4 -- 1.4&-2.2&0.20&8.8E-01\\
0626+24&0.4 -- 4.9&-1.6&0.08&1.2E-03&&1718$-$02&0.4 -- 1.4&-2.2&0.16&1.6E-06\\
0628$-$28&0.4 -- 10.6&-1.9&0.10&6.8E-01&&1718$-$32&0.4 -- 1.4&-2.3&0.06&3.9E-01\\
0643+80&0.4 -- 4.9&-1.9&0.08&2.3E-01&&1726$-$00&0.4 -- 0.6&-2.3&0.47&\\
0655+64&0.4 -- 1.4&-2.1&0.30&1.0E-01&&1727$-$33&0.4 -- 1.4&-1.3&&\\
0656+14&0.4 -- 1.4&-0.5&0.17&1.3E-01&&1730$-$22&0.4 -- 1.4&-2.0&0.15&1.4E-01\\
0727$-$18&0.4 -- 1.4&-1.6&0.11&1.3E-02&&1732$-$07&0.4 -- 1.4&-1.9&0.12&1.4E-09\\
0740$-$28&0.4 -- 10.6&-2.0&0.03&1.1E-07&&1734$-$35&0.6 -- 1.4&-1.6&0.30&\\
0751+32&0.4 -- 4.9&-1.5&0.07&1.8E-01&&1735$-$32&0.4 -- 1.6&-0.9&0.12&2.1E-02\\
0756$-$15&0.4 -- 4.9&-1.6&0.13&6.0E-04&&1736$-$31&0.6 -- 1.6&-0.9 ~~~~~&0.20&4.2E-01\\
0809+74&0.4 -- 10.6&-1.4&0.06&6.7E-02&&1737+13&0.4 -- 4.9&-1.5&0.10&2.8E-01\\
0818$-$13&0.4 -- 4.9&-2.3&0.05&3.4E-01&&1737$-$30&0.4 -- 1.4&-1.3&0.10&4.5E-01\\
0820+02&0.4 -- 4.9&-2.4&0.08&9.5E-01&&1738$-$08&0.4 -- 4.9&-2.2&0.08&6.9E-02\\
0834+06&0.4 -- 4.9&-2.7&0.11&2.1E-02&&1740$-$03&0.4 -- 1.4&-1.5&&\\
0853$-$33&0.4 -- 1.4&-2.4&0.20&5.4E-01&&1740$-$13&0.4 -- 1.4&-2.0&0.19&2.3E-02\\
0906$-$17&0.4 -- 1.4&-1.4&0.16&2.0E-01&&1740$-$31&0.6 -- 1.4&-1.9&0.11&\\
0917+63&0.4 -- 1.4&-1.7&0.37&&&1742$-$30&0.4 -- 4.9&-1.6&0.04&1.5E-06\\
0919+06&0.4 -- 10.6&-1.8&0.05&7.6E-01&&1745$-$12&0.4 -- 1.4&-2.1&0.12&3.4E-03\\
0940+16&0.4 -- 1.4&-1.3&0.30&5.3E-01&&1746$-$30&0.4 -- 1.4&-1.5&0.39&\\
0942$-$13&0.4 -- 1.4&-3.0&0.30&8.8E-01&&1747$-$31&0.6 -- 1.4&-1.2&0.31&\\
1750$-$24&0.9 -- 4.9&-1.0&0.07&1.1E-06&&1838$-$04&0.9 -- 1.6&-1.3&0.21&2.0E-01\\
\hline
\end{tabular}
\end{table*}

\setcounter{table}{1}
\begin{table*}
\caption{- continued }
\begin{tabular}{lllllllllll}
\hline
PSR B & Freq. range & $\alpha$ & $\sigma_\alpha$ & Q& &PSR B & Freq. range & $\alpha$ & $\sigma_\alpha$ & Q \\
 & [GHz] & & & \\
\hline
1753+52&0.4 -- 4.9&-1.6&0.08&5.8E-04&&1839+09&0.4 -- 1.4&-2.0&0.07&3.1E-01\\
1753$-$24&0.4 -- 1.6&-0.7&0.14&3.2E-01&&1839+56&0.4 -- 1.4&-1.5&0.22&3.2E-01\\
1754$-$24&0.4 -- 1.4&-1.1&0.09&9.1E-02&&1839$-$04&0.4 -- 1.6&-1.6&0.08&1.7E-01\\
1756$-$22&0.4 -- 4.9&-1.7&0.09&5.3E-01&&1841$-$04&0.4 -- 1.6&-1.6&0.07&5.8E-03\\
1757$-$24&0.4 -- 0.6&-3.6&&&&1841$-$05&0.6 -- 4.9&-1.7&0.10&5.3E-01\\
1758$-$03&0.4 -- 1.4&-2.6&0.11&2.3E-01&&1842+14&0.4 -- 4.9&-1.6&0.09&1.7E-02\\
1758$-$23&1.4 -- 4.9&-2.5&0.10&8.3E-04&&1842$-$02&0.6 -- 1.6&-0.9&0.27&4.1E-02\\
1800$-$27&0.4 -- 1.4&-1.4&&&&1842$-$04&0.6 -- 1.4&-0.8&0.29&1.4E-01\\
1802$-$07&0.4 -- 1.4&-1.3&0.31&&&1844$-$04&0.4 -- 4.9&-2.2&0.06&8.2E-02\\
1804$-$08&0.4 -- 4.9&-1.2&0.08&6.1E-07&&1845$-$01&0.4 -- 10.6&-1.6&0.05&4.0E-01\\
1804$-$27&0.6 -- 1.4&-3.0&0.21&&&1845$-$19&0.4 -- 0.6&-2.0&0.46&\\
1805$-$20&0.6 -- 4.9&-1.5&0.07&&&1846$-$06&0.4 -- 1.4&-2.2&0.10&8.4E-01\\
1806$-$21&0.6 -- 1.6&-2.0&0.34&7.5E-02&&1848+04&0.6 -- 1.4&-1.4&&\\
1810+02&0.4 -- 1.4&-1.7&0.21&3.1E-02&&1848+12&0.4 -- 1.6&-1.9&0.16&4.3E-02\\
1811+40&0.4 -- 1.4&-1.8&0.22&5.0E-02&&1848+13&0.4 -- 1.6&-1.4&0.18&1.2E-01\\
1813$-$17&0.6 -- 1.6&-1.0&0.14&5.0E-01&&1849+00&1.4 -- 4.9&-2.4&0.12&4.8E-01\\
1813$-$26&0.4 -- 0.6&-1.4&0.33&&&1851$-$14&0.4 -- 0.6&-0.8&0.42&\\
1815$-$14&0.9 -- 1.6&-1.6&0.22&8.5E-04&&1853+01&0.4 -- 0.6&-2.5&&\\
1817$-$13&0.6 -- 1.6&-1.7&0.37&3.7E-01&&1855+02&0.6 -- 4.9&-1.2&0.09&9.5E-02\\
1817$-$18&0.4 -- 1.4&-1.1&0.27&&&1857$-$26&0.4 -- 10.7&-2.1&0.06&1.2E-03\\
1818$-$04&0.4 -- 4.9&-2.4&0.06&9.7E-02&&1859+01&0.4 -- 0.6&-2.9&0.21&\\
1819$-$22&0.4 -- 1.4&-1.7&0.07&2.1E-01&&1859+03&0.4 -- 4.9&-2.8&0.08&3.8E-02\\
1820$-$11&0.4 -- 4.9&-1.5&0.05&1.9E-04&&1859+07&0.4 -- 1.6&-1.0&0.15&8.2E-01\\
1820$-$14&0.6 -- 1.4&-0.7&0.22&&&1900+01&0.4 -- 4.9&-1.9&0.15&3.3E-01\\
1820$-$30B&0.4 -- 0.6&-1.9&0.33&&&1900+05&0.4 -- 4.9&-1.7&0.08&3.9E-02\\
1820$-$31&0.4 -- 1.4&-2.1&0.20&4.8E-01&&1900+06&0.4 -- 4.9&-2.2&0.10&1.4E-01\\
1821+05&0.4 -- 1.4&-1.7&0.18&2.1E-02&&1900$-$06&0.4 -- 0.6&-1.8&0.19&\\
1821$-$11&0.6 -- 4.9&-2.3&0.10&1.2E-01&&1902$-$01&0.4 -- 1.4&-1.9&0.11&1.1E-02\\
1821$-$19&0.4 -- 4.9&-1.9&0.06&2.2E-01&&1903+07&0.6 -- 1.4&-1.3&0.10&2.6E-01\\
1822+00&0.4 -- 1.4&-2.4&0.26&4.0E-01&&1904+06&0.4 -- 1.6&-0.7&0.21&9.6E-01\\
1822$-$09&0.4 -- 10.6&-1.3&0.08&1.3E-02&&1905+39&0.4 -- 1.4&-2.0&0.16&9.0E-02\\
1822$-$14&1.4 -- 4.9&-0.7&0.08&3.6E-01&&1907+00&0.4 -- 1.4&-2.0&0.11&5.6E-08\\
1823$-$11&0.4 -- 1.6&-2.4&0.10&9.9E-01&&1907+02&0.4 -- 1.4&-2.8&0.11&3.8E-02\\
1823$-$13&0.6 -- 10.6&-0.6&&&&1907+03&0.4 -- 4.9&-1.8&0.07&1.4E-01\\
1826$-$17&0.4 -- 4.9&-1.7&0.06&5.1E-05&&1907+10&0.4 -- 1.4&-2.5&0.09&5.6E-01\\
1828$-$10&0.4 -- 1.6&-0.4&0.14&2.3E-01&&1907$-$03&0.4 -- 1.4&-2.7&0.12&7.4E-05\\
1829$-$08&0.4 -- 1.6&-0.8&0.06&1.5E-04&&1910+20&0.4 -- 1.4&-1.6&0.16&5.9E-01\\
1829$-$10&0.4 -- 1.6&-1.3&0.15&1.4E-01&&1911+11&0.4 -- 0.6&-1.4&&\\
1830$-$08&0.6 -- 4.9&-1.1&0.05&1.2E-21&&1911+13&0.4 -- 4.9&-1.5&0.06&8.5E-03\\
1831$-$00&0.4 -- 0.6&-1.4&&&&1911$-$04&0.4 -- 1.4&-2.6&0.11&3.3E-01\\
1831$-$03&0.4 -- 1.4&-2.7 ~~~~~&0.08&1.1E-04&&1913+10&0.4 -- 1.6&-1.9&0.15&5.6E-01\\
1831$-$04&0.4 -- 4.9&-1.3&0.07&9.5E-01&&1913+16&0.4 -- 1.4&-1.4&0.24&\\
1832$-$06&0.6 -- 1.6&-0.4&0.35&2.9E-01&&1913+167&0.4 -- 0.6&-1.4&0.59&\\
1834$-$04&0.6 -- 1.6&-1.9&0.30&5.8E-01&&1914+09&0.4 -- 1.4&-2.3&0.11&2.1E-03\\
1834$-$06&0.6 -- 1.6&-1.2&0.24&9.5E-01&&1914+13&0.4 -- 4.9&-1.6&0.10&3.0E-02\\
1834$-$10&0.4 -- 1.6&-2.1&0.09&3.1E-01&&1915+13&0.4 -- 4.9&-1.8&0.09&3.7E-02\\
1916+14&0.4 -- 1.4&-0.3&0.43&1.0E+00& &2027+37&0.4 -- 1.4&-2.5&0.10&4.6E-02\\
1917+00&0.4 -- 4.9&-2.2&0.07&6.3E-01& &2035+36&0.4 -- 1.7&-1.6&0.39&9.8E-01\\
1918+19&0.4 -- 1.4&-2.4&0.16&9.4E-01& &2036+53&0.4 -- 1.4&-2.0&0.27&2.2E-01\\
1919+14&0.4 -- 4.9&-1.3&0.14&9.2E-01& &2044+15&0.4 -- 1.4&-1.7&0.15&3.4E-03\\
1919+21&0.4 -- 4.9&-2.6&0.04&0.0E+00& &2045+56&0.4 -- 1.4&-2.4&&\\
1920+20&0.4 -- 0.6&-2.5&0.70&& &2045$-$16&0.4 -- 10.6&-2.1&0.07&1.8E-01\\
1920+21&0.4 -- 4.9&-2.2&0.07 &2.8E-02& &2053+21&0.4 -- 1.4&-0.8&0.35&\\
1923+04&0.4 -- 0.6&-2.7&0.50& & &2053+36&0.4 -- 1.4&-1.9&0.04&3.5E-03\\
1924+16&0.4 -- 1.4&-1.5&0.16&2.9E-01& &2106+44&0.4 -- 4.9&-1.4&0.06&7.5E-04\\
1929+10&0.4 -- 24&-1.6&0.04&1.5E-07& &2110+27&0.4 -- 1.4&-2.2&0.18&5.6E-01\\
\hline
\end{tabular}
\end{table*}

\setcounter{table}{1}
\begin{table*}
\caption{- continued }
\begin{tabular}{lllllllllll}
\hline
PSR B & Freq. range & $\alpha$ & $\sigma_\alpha$ & Q& &PSR B & Freq. range & $\alpha$ & $\sigma_\alpha$ & Q \\
 & [GHz] & & & \\
\hline
1929+20&0.4 -- 1.4&-2.5&0.22&7.4E-01& &2111+46&0.4 -- 10.6&-2.1&0.04&2.1E-02\\
1930+22&0.4 -- 1.6&-1.5&0.09&3.9E-02& &2113+14&0.4 -- 1.4&-1.9&0.13&3.7E-02\\
1931+24&0.4 -- 0.6&-4.0&&& &2148+52&0.4 -- 1.6&-1.3&0.05&1.1E-04\\
1933+16&0.4 -- 4.9&-1.7&0.03&8.1E-03& &2148+63&0.4 -- 2.7&-1.8&0.09&1.1E-03\\
1935+25&0.4 -- 1.4&-0.7&0.20&5.6E-02& &2152$-$31&0.4 -- 1.4&-2.3&0.40&3.9E-01\\
1937$-$26&0.4 -- 1.4&-0.9&0.30&1.9E-01& &2154+40&0.4 -- 4.9&-1.6&0.09&3.6E-02\\
1940$-$12&0.4 -- 1.4&-2.4&0.20&5.1E-01& &2210+29&0.4 -- 1.4&-1.5&0.20&1.3E-01\\
1941$-$17&0.4 -- 1.4&-2.3&0.30&& &2217+47&0.4 -- 4.9&-2.6&0.19&4.5E-01\\
1942$-$00&0.4 -- 1.4&-1.8&0.17&2.5E-02& &2224+65&0.4 -- 4.6&-1.9&0.11&3.9E-01\\
1943$-$29&0.4 -- 1.4&-2.0&0.25&8.9E-01& &2227+61&0.4 -- 1.4&-2.6&0.10&5.7E-03\\
1944+17&0.4 -- 4.9&-1.3&0.12&1.9E-01& &2241+69&0.4 -- 1.4&-1.4&0.46&5.8E-01\\
1946+35&0.4 -- 4.9&-2.4&0.04&6.9E-09& &2255+58&0.4 -- 4.9&-2.1&0.07&1.0E+00\\
1946$-$25&0.4 -- 1.4&-2.0&0.22&5.4E-01& &2303+30&0.4 -- 1.4&-2.3&0.16&3.6E-03\\
1951+32&0.4 -- 1.6&-1.6&0.11&5.2E-03& &2303+46&0.4 -- 1.6&-1.6&&\\
1953+50&0.4 -- 4.9&-1.6&0.09&8.5E-01& &2306+55&0.4 -- 4.9&-1.9&0.06&3.1E-01\\
2000+32&0.4 -- 4.9&-1.1&0.04&3.1E-02& &2310+42&0.4 -- 10.6&-1.9&0.03&5.1E-05\\
2000+40&0.4 -- 4.9&-2.2&0.03&5.5E-30& &2315+21&0.4 -- 1.4&-2.1&0.44&6.9E-01\\
2002+31&0.4 -- 1.4&-1.7&0.06&2.3E-20& &2323+63&0.4 -- 1.4&-0.8&0.23&6.4E-02\\
2003$-$08&0.4 -- 1.4&-1.4&0.20&9.4E-02& &2327$-$20&0.4 -- 1.4&-2.0&0.29&8.7E-01\\
2016+28&0.4 -- 10.6&-2.2&0.04&4.7E-01& &2334+61&0.4 -- 1.4&-1.7&0.23&6.3E-01\\
2022+50&0.4 -- 4.9&-0.8&0.05&5.3E-01& &2351+61&0.4 -- 10.6&-1.1&0.13&3.1E-01\\
\hline
\end{tabular}
\end{table*}

\begin{table*}
\caption{Spectral indices for 15 pulsars with two-power-law spectrum calculated for a given frequency range with error and a~probability of the goodness-of-fit Q. The break frequency $\nu_b$ is indicated.}
\begin{tabular}{llllllllll}
\hline
PSR B & Freq. range & $\alpha_1$ & $\sigma_{\alpha1}$ & $Q_1$ &  $\alpha_2$ & $\sigma_{\alpha2} $ & $Q_2$ & $\nu_b$\\
 & [GHz] & & & & & & & [GHz] &\\
\hline
&&&\\
0136+57&0.4 -- 4.9&-1.1&0.13&8.7E-02&-2.3&0.35&1.3E-02&1.0\\
0226+70&0.4 -- 1.4&-0.5&0.24&5.8E-01&-4.0&0.85&&0.9\\
0301+19&0.4 -- 4.9&-0.9&0.38&5.0E-01&-2.3&0.34&&0.9\\
0355+54&0.4 -- 23.0&-0.7&0.19&1.0E-02&-1.2&0.04&3.9E-01&1.9\\
0540+23&0.4 -- 32.0&-0.3&0.14&8.3E-01&-1.6&0.09&6.0E-05&1.4\\
0823+26&0.4 -- 14.8&-0.7&0.43&4.5E-01&-2.1&0.08&3.1E-05&1.3\\
1237+25&0.4 -- 10.7&-0.9&0.19&7.4E-08&-2.2&0.25&2.5E-04&1.4\\
1749$-$28&0.4 -- 10.7&-2.4&0.06&3.8E-02&-4.3&0.36&1.5E-01&2.7\\
1800$-$21&0.4 -- 4.9&-0.2&0.07&1.3E-01&-1.0&0.32&&1.4\\
1952+29&0.4 -- 10.7&-0.6&0.52&9.2E-01&-2.7&0.10&6.8E-01&1.2\\
2011+38&0.4 -- 4.9&-0.9&0.13&1.5E-01&-1.9&0.10&7.3E-03&1.4\\
2020+28&0.4 -- 32.0&-0.7&0.40&3.3E-01&-1.9&0.17&6.9E-01&2.3\\
2021+51&0.4 -- 23.0&-0.8&0.20&1.8E-01&-1.5&0.07&2.0E-01&2.6\\
2319+60&0.4 -- 10.6&-1.1&0.12&7.3E-07&-2.1&0.05&6.6E-03&1.4\\
2324+60&0.4 -- 4.9&-1.2&0.12&5.8E-01&-2.5&0.27&&1.4\\
\hline
\end{tabular}
\end{table*}

We found only 15 pulsars out of 167 whose spectral fit evidently required the two-power-law model (Table 3). In Fig.~\ref{1}a we
present distribution of spectral indices $\alpha$ for pulsars with a simple power-law
spectrum and in Fig.~\ref{1}b and c for pulsars with a broken-type spectrum (see caption for explanation of $\alpha_1$ and $\alpha_2$). In Fig.~\ref{2}b and c we also show two examples of two-power-law spectrum pulsars with the smallest and largest slope difference, respectively.

As detailed above, in order to reduce the effects of diffractive and
refractive interstellar scintillations as well as possible intrinsic
phenomena, we need a large number of measurements at a given frequency
to obtain reliable pulsar spectra. We believe that our large sample of
flux density measurements is capable of doing this over a wide
frequency range, allowing an analysis of the spectral behaviour of
pulsar radio emission. Our analysis shows that, in principle, pulsar
spectra are described by a simple power law with the mean spectral
index $<\alpha>=-1.8\pm0.2$ (see Fig.~\ref{2}a). We examined the data for
any possible correlations between spectral index and rotation period $P$, spin-down
rate $\dot P$, characteristic age $\tau$, polarization as well as profile
type. In general no significant correlations were found but we
have distinguished some interesting groups of objects which are
discussed below:\\

(i) {\it Very steep spectrum sources} \\
This group of pulsars consists of objects
with very steep spectra. Examples of such pulsars are the PSRs
B0942$-$13, B0943+10 and B1859+03$^1$ with spectral indices of $-$3.0,
$-$3.7 and $-$2.8, respectively (see Table 2). Lorimer et
al. (\cite{lorimer}) suggested that older pulsars ($\tau \geq 10^8$
yr) have steeper spectra, which is obviously not the case for B0943+10 and
B1859+03 as these pulsars have characteristic ages of $4.9 \times
10^6$ yr and $1.4 \times 10^6$~yr, respectively. There is, of course,
yet another
exception, the Crab pulsar (PSR B0531+21), i.e.~the youngest known
radio pulsar with
one of the steepest spectrum in our sample. These results provide
evidence that
there is no correlation between steepness of spectra and the characteristic pulsar age.\\

(ii) {\it Flat spectrum sources}\\
It was previously believed that pulsars have a steep spectra but the analysis of a large sample shows that there are pulsars with flat spectra over a wide frequency range. In this group there are pulsars which have almost flat spectra with $\alpha
\geq -1.0$. Examples of such pulsars are B0144+59,
B1750$-$24 and B2022+50$^1$ with spectral indices of $-1.0\pm 0.04$, $-1.0\pm 0.07$ and
$-0.8\pm 0.05$, respectively. Although Lorimer et
al. (\cite{lorimer}) suggested that younger pulsars have flat spectra,
they also found that PSR B1952+29 possessed a flat spectrum and yet
had a characteristic age of $3.4 \times 10^9$ yr. This pulsar indeed
has a flat spectrum in the frequency range from 400 MHz to 1.4 GHz but
considering the whole frequency range from 400 MHz to 10.7 GHz its
spectrum becomes a two-power-law one. A similar behaviour was observed
for PSR B0540+23$^1$. It is possible that the pulsars with flat spectrum
mentioned by Lorimer et al. (\cite{lorimer}) may have a break in
their spectrum at higher frequencies.\\

\begin{figure}
\setlength{\unitlength}{1cm}
\begin{picture}(1,10)
\put(-1,10){\includegraphics{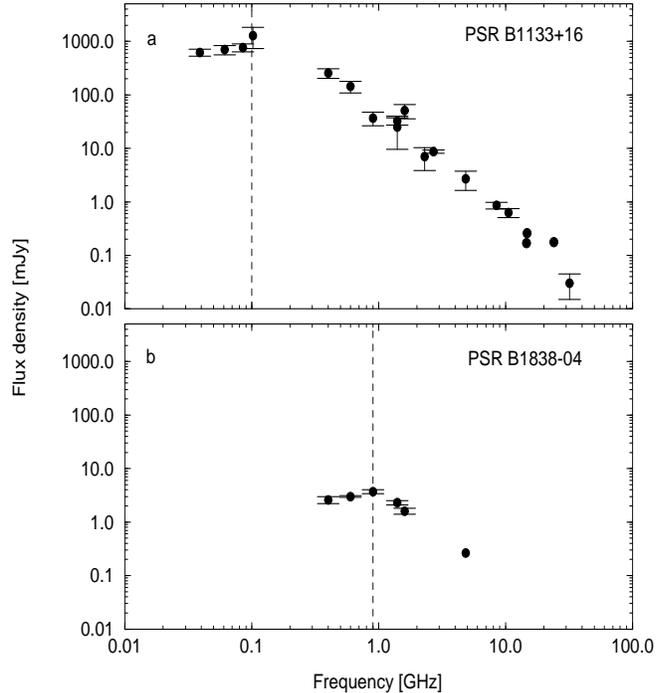}}
\end{picture}
\caption{{\bf a} Typical low frequency turn-over,
{\bf b} an example of the unusual turn-over at around 1~GHz.
}
\label{3}
\end{figure}

(iii) {\it Sources with low-frequency turn-over}\\ 
Spectra of many pulsars show a low-frequency turn-over at $\sim$ 100
MHz (Sieber \cite{sieber73}, Izvekova et al. \cite{izvekova}). We have not fitted the
turn-over points because of the gap in flux density measurements at
frequencies between 100 MHz and 300 MHz and difficulties in
determining the maximum frequency $\nu_{{\rm max}}$. We have found 2
pulsars in our sample which have a turn-over at unusually high
frequency ($\sim$ 1 GHz): B1838$-$04 and B1823$-$13 (see Fig.~\ref{3}). These are young
pulsars and all belong to the 1800$-$21-class of pulsars, which was
introduced by Hoensbroech et al. (\cite{hoensbroech98}) to describe
their unusual polarization properties. The PSR B1800$-$21$^1$ has a two power law spectrum, which may be also interpreted as a ``broad form of 
turn-over''.\\

(iv) {\it Sources with high-frequency turn-up or flattening}\\
There is a group of pulsars which have a possible turn-up or
flattening in spectra at very high frequencies. Therefore we have not
fitted the points above 23 GHz. In this group there are pulsars such
as: B0329+54, B0355+54, B1929+10 and B2021+51$^1$ which were studied in
detail by Kramer et al. (\cite{kramer96}). There are also pulsars
which may show a spectrum flattening already at $\sim$ 5 GHz. For
example, the spectra of PSR B0144+59 and B2255+58$^1$. Recently, the idea of a spectral
change at very high frequencies received a strong support from
observations of the Crab pulsar. Moffet \& Hankins
(\cite{moffet}) observed a clear flattening of its spectrum at realtively low frequency around 10 GHz, as compared with about 20 GHz (Kramer et al. \cite{kramer96}) .\\

(v) {\it Sources with two power law spectra}\\
We also recognized broken-type spectra (Sieber \cite{sieber73} and Malofeev et al. \cite{malofeev94}), although
there are only 15 definite two-power-law cases in our sample, showing 
so called break frequency between 0.9 and 2.7 GHz (see Table 3) which divides the whole spectrum into two parts with considerably different slopes. This is
a significantly smaller fraction (only about 10\%) than the reported 35\% by Malofeev
(\cite{malofeev96}). The distribution of spectral indices for broken-type
pulsars is shown in Fig.~\ref{1}b and c.  The reduced fraction of two-power-law
spectra pulsars can be only partly explained by selection effects
possibly present in the Malofeev sample. In fact, many pulsars which
were previously thought to demonstrate two-power-law spectra (Malofeev et
al. \cite{malofeev94}, Kramer et al. \cite{kramer96}, Xilouris et
al. \cite{xilouris}) can be modelled by a simple power-law spectra in our sample. We
believe that this can be largely explained by severe flux density
variations (see Sect. 2.2) and the fact that the number of
measurements included into a fit so far  may have been too small
(e.g.~PSR B0628+28$^1$). 

We note that Gil et al. (\cite{gil94}) and Malofeev
(\cite{malofeev96}) suggested that PSR B1822$-$09 has a complex flux
density spectrum. Our data do not confirm such a behaviour although
some deviations from a simple power law seem to be present.  Moreover,
there are indeed some other pulsars which exhibit a rather unusual
spectral behaviour (e.g. B0823+26, B0621$-$04, B0656+14$^1$). These objects may
actually have a complex spectrum but we certainly need more and better
data before we should consider them as a separate class of sources.
Generally, pulsar spectra can be classified in two groups: Out of
167 pulsars which have been studied over a wide frequency range (from
400 MHz up to at least 5 GHz), the spectra of most pulsars can be
modelled by a simple power law. About 10\% of all pulsars require two
power laws to fit the data. Table 4 and references therein clearly indicate that the spectra of slowly and
fast rotating pulsars (i.e.~millisecond pulsars) are indeed identical
on the average. 
We note again, that the analysis of the main physical
parameters of pulsars with unusual or two-power-law spectra has not
shown any correlation, consistent with the results of Xilouris et
al. (\cite{xilouris}) and Malofeev (\cite{malofeev96}).

\section{Summary and Conclusions}

The main conclusion of this paper is that the single power law spectrum is a rule and the two power law spectrum is a rare exception to this rule. One could therefore think that the nature of this exception is that the spectrum just becomes steeper starting from some, relatively high frequency. However, inspection of Fig.~\ref{1} indicates that this might not be a case. The distribution of $\alpha_1$ seems different from that of $\alpha$, meaning that in pulsars with two power law spectra the average spectral index $<\alpha_1>$ is typically much larger than the average $<\alpha>$ (Fig.~\ref{1}b) and of course, the high frequency index $<\alpha_2>$ is much smaller than $<\alpha>$ (Fig.~\ref{1}c). Thus, it seems that the two power law spectra are qualitatively different from the typical single power law spectra.
In this paper we obtained spectral index for a large sample of pulsars
in a wide frequency range (form 400 MHz to 23 GHz) . The average
spectral index of pulsars $<\alpha>$ with simple power-law spectrum in
our sample is $\rm-1.8\pm0.2$ which agrees with results obtained by
other authors (see Table 4).  The distribution of spectral indices is
symmetric and almost Gaussian. The average indices for the
broken-type spectra are $<\alpha_1> = -0.9\pm0.5$ and $<\alpha_2> =
-2.2\pm0.9$, respectively, with a break frequency of $<{\nu}_b>
\approx 1.5$ GHz on the average.  We have not found any
correlations between spectral index and rotation period $P$, spin-down
rate $\dot P$, characteristic age $\tau$, polarization and profile
type for pulsars with both simple power law spectra and
two-power-law spectra. We have found 2 young, nearly fully polarized pulsars which indicate turn-over at unusually high frequency ($\sim$~1~GHz). We have
also found 15 pulsars which definitely require two-power-law spectra.
The comparison of pulsar spectra analysis for slow and millisecond
pulsars indicates that both groups have the same emission mechanism
(Table~4).

\begin{table}
\caption{Spectral indices obtained by different authors.}
\begin{tabular}{crcl}
\hline
Spectral &	No. of & Freq. range & References\\
index & PSRs & [GHz]  &\\
\hline
-1.6 &	  27 &	0.1 -- 10 &	Sieber 1973\\
-1.9 &        20 &    0.1 -- 30          &               Malofeev et al. 1994\\
-1.6 &	280 &	0.3 -- 1.6 &	Lorimer et al. 1995\\
-1.7 &	284&	0.1 -- 10	&	Malofeev et al. 1996\\
-1.8 &	  32 &	0.3 -- 4.9 &	Kramer et al. 1998, 1999\\
       &           &                         &              (millisecond PSRs)\\
-1.7 &	216 &	0.4 -- 1.5 &	Toscano et al. 1998\\
      &           &                         &               (southern PSRs)\\
-1.9 &	19 &	0.4 -- 1.5 &	Toscano et al. 1998\\
      &           &                         &               (millisecond PSRs)\\
-1.9 &	144 &	1.4 -- 4.9 &	Kijak et al. 1998\\
-1.8 &	281 &	0.4 -- 23 &	this paper\\
\hline
\end{tabular}
\end{table}

Postscript files of spectra are available at http://astro.ca.wsp.zgora.pl/olaf/paper1 and our spectra are also presented in EPN Database at http://www.mpifr-bonn.mpg.de

\begin{acknowledgements}
We thank Christoph Lange for his help with our observations in August 1998. OM thanks the director of the MPIfR Prof. Dr. R. Wielebinski for invitation and support. OM and JK gratefully acknowledge several discussions with J. Gil in the course of this work. The authors also thank V.M. Malofeev for unpublished data at 102 MHz and his helpful comments. This work was supported in part by the Polish State Committee for Scientific Research Grant 2~P03D~008~19
\end{acknowledgements}

\end{document}